\documentclass[twocolumn,prl,superscriptaddress,floatfix,aps]{revtex4}
\newcommand{\be}{\begin{equation}}
\newcommand{\ee}{\end{equation}}

\usepackage{graphicx}
\usepackage{dcolumn}
\usepackage{bm}
\usepackage{psfig}
\usepackage{longtable}

\begin{document}
\title{Multi-scale coarse-graining of diblock copolymer self-assembly: from monomers to ordered micelles}
\author{Carlo Pierleoni}
\affiliation{INFM CRS-SOFT, and Department of Physics, U. of L'Aquila, Via
Vetoio, I-67010 L'Aquila, Italy}
\author{Chris Addison}
\affiliation{Dept. of Chemistry, University of Cambridge, Cambridge CB2 1EW, United Kingdom}
\author{Jean-Pierre Hansen}
\affiliation{Dept. of Chemistry, University of Cambridge, Cambridge CB2 1EW, United Kingdom}
\author{Vincent Krakoviack}
\affiliation{Laboratoire de Chimie, Ecole Normale Sup\'erieure de Lyon, 69364 Lyon Cedex 07,
France}


\begin{abstract}
Starting from a microscopic lattice model, we investigate clustering, micellization and micelle ordering in semi-dilute solutions of AB diblock copolymers in a selective solvent. To bridge the gap in length scales, from monomers to ordered micellar structures, we implement a two-step coarse graining strategy, whereby the AB copolymers are mapped onto ``ultrasoft'' dumbells with monomer-averaged effective interactions between the centres of mass of the blocks. Monte Carlo simulations of this coarse-grained model yield clear-cut evidence for self-assembly into micelles with a mean aggregation number $n\simeq 100$ beyond a critical concentration. At a slightly higher concentration the micelles spontaneously undergo a disorder-order transition to a cubic phase. We determine the effective potential between these micelles from first principles.
\end{abstract}

\pacs{PACS:  }
\maketitle

Block copolymers in solution show a remarkable tendency to self-assemble into a bewildering range of disordered (liquid-like) and ordered structures, depending on the macromolecular composition, the relative sizes of the blocks, solvent selectivity, polymer concentration and temperature \cite{Lodge05,Hamley}. Complex phase diagrams of styrene-isoprene diblock copolymers in various solvents have been mapped out experimentally in great detail by varying the above physical parameters, using a combination of light scattering, small-angle neutron and X-ray scattering and rheological studies \cite{Lodge02,Park05}. A frequent, partial scenario sees the copolymers aggregate into polydisperse spherical micelles at fairly low polymer concentration; upon lowering the temperature or increasing the concentration, the micelles undergo a disorder-order transition onto a cubic lattice. A theoretical understanding of copolymer phase behaviour generally relies on phenomenological considerations \cite{Zhulina05}, or on self-consistent field theory similar to that applied earlier to copolymer melts \cite{Leibler80}. A more molecular approach is obviously very difficult, in view of the wide range of length  scales involved, from the monomer level to the mesoscopic scales characterising ordered micellar structures. In particular, monomer-level Monte Carlo (MC) simulations have so far been restricted to short copolymer chains \cite{Milchev01}.

In this letter we propose and implement a two-stage coarse graining strategy to investigate the micellization and disorder-order transition of a simple microscopic model of diblock copolymers. The AB copolymers are made up of two blocks of equal numbers $M_{A}=M_{B}=M$ of monomers on a cubic lattice. $N$ such chains occupy a fraction of the lattice sites. The remaining sites are occupied by an implicit solvent chosen to be selective (good solvent) for the B blocks, while it is a $\theta$-solvent for the A blocks. This situation is crudely represented by modelling the A blocks as ideal chains (I) and the B blocks as self and mutually avoiding walks (S). Moreover A and B blocks of the same or different copolymers are assumed to be mutually avoiding (S). The resulting model will be referred to as ISS. The model is clearly athermal, so that temperature will be irrelevant, and the polymer density $\rho=N/L^{3}$ 
is the only thermodynamic variable ($L^{3}$ being the total number of sites of the cubic lattice, the spacing $a$ of which will be taken as the unit of length). With typical polymer sizes $M\simeq 10^{3}$, and expected micelle aggregation numbers $n\simeq 10^{2}$, very large system sizes would be needed to generate the tens of micelles required to study their ordered and disordered structures. 

The first stage of our coarse-graining procedure, designed to overcome the above bottleneck, is inspired by earlier work on homopolymers \cite{Dautenhahn94, Bolhuis01}. AB copolymers are mapped onto ``ultrasoft'' asymmetric dumbells with effective interactions $v_{\alpha\beta}(R)$ between the centres of mass (CM) of A and B blocks on different copolymers, and an intramolecular ``tethering'' potential $\phi_{AB}(R)$ between the CM's of the two blocks of the same copolymer \cite{Addison05}. Since the original model is athermal, these effective interactions are purely entropic and are determined by taking statistical averages over copolymer conformations for fixed A-A, A-B and B-B CM distances $R$. In practice this is achieved by inverting the CM-CM intermolecular and intramolecular pair distribution functions $g_{AA}(R), g_{AB}(R), g_{BB}(R)$ and $s_{AB}(R)$. The inversion was carried out in the low density limit, using pair distribution functions generated in monomer-level MC simulations of two diblock copolymers with $2M=500$ monomers each, and an exact relation between the $g_{\alpha\beta}(R), s_{AB}(R)$ and the corresponding $v_{\alpha\beta}(R), \phi_{AB}(R)$ \cite{Addison05}. 
\begin{figure}[htb]
\centering
\includegraphics[scale=0.35]{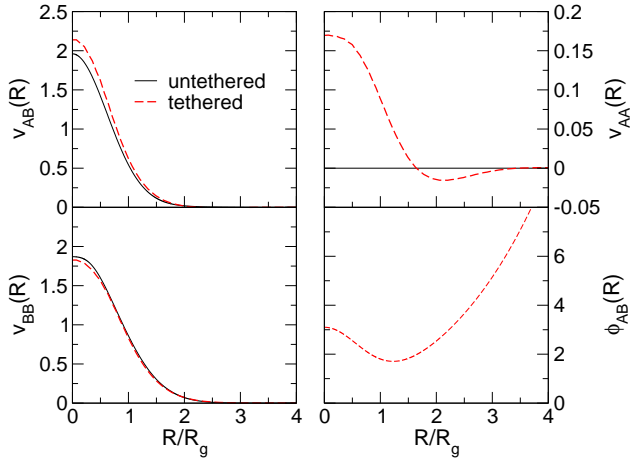}
\caption{(Colour on line) Effective pair potentials $v_{AA}(R), v_{AB}(R), v_{BB}(R)$ and  $\phi_{AB}(R)$ for tethered and untethered block copolymers in the zero density limit. }
\label{fig:fig1}
\end{figure}
The resulting pair potentials are shown in figure \ref{fig:fig1}, where they are compared to the corresponding pair potentials for an equimolar binary mixture of untethered A and B homopolymers; in the latter case the inversion procedure is trivial in the zero density limit where the effective potentials $v_{\alpha\beta}(R)$ and $\phi_{AB}(R)$ reduce to the potentials of mean force, e.g. $v_{\alpha\beta}(R)=-k_{B}T \ln g_{\alpha\beta}(R)$. The two sets of effective pair potentials $v_{AB}(R)$ and $v_{BB}(R)$ are seen to be surprisingly close. As in the case of homopolymers \cite{Dautenhahn94, Bolhuis01} they are roughly gaussian in shape with an amplitude of a few $k_{B}T$, and a range of the order of the natural length scale, namely the radius of gyration $R_{g}$ of the copolymer. In lattice units, the MC simulations yield the following radii for an isolated copolymer with $2M=500$: $R_{g,A}=6.964, R_{g,B}=11.274, R_{g}=13.889$, while the root mean square distance between the CM's of the two blocks is $R_{AB}=20.661$. The results for $v_{AB}(R)$ and $v_{BB}(R)$ in figure \ref{fig:fig1} suggest that tethering of the two blocks does not significantly modify their effective interactions. This is no longer true for $v_{AA}(R)$, since this potential is strictly zero in the untethered (binary mixture) case, while it is about $10\%$ of $v_{BB}(R)$ for the copolymer; this residual effective interaction between the ideal blocks is induced by the tethering to non-ideal B-blocks. 

The key assumption made in this work is that these zero density pair potentials are transferrable to finite copolymer densities, i.e. we neglect the density dependence of the effective potentials. Earlier experience with homopolymers shows that this is not an unreasonable approximation well into the semi-dilute regime $\rho/\rho^{*} \lesssim 5$ where $\rho^{*}=(4\pi R_{g}^{3}/3)^{-1}$ is the overlap density \cite{Bolhuis01}. 

We have carried out extensive MC simulations of samples of $5\times10^{3}$ and $10^{4}$ AB copolymers using the effective pair potentials of figure \ref{fig:fig1}, in the range $3\leq\rho/\rho^{*}\leq 10$. The simulations using the coarse-grained representation are at least two order of magnitude faster than simulations of the original all-monomer lattice model. This allowed us to make detailed investigations, using both sets of potentials in figure \ref{fig:fig1}, in order to ascertain the sensitivity of the micellization behaviour to details of the effective pair potentials. Both sets of potentials yield results in semi-quantitative agreement, and only those obtained with the untethered potentials will be shown here. Inspection of snapshots of configurations shows that micellization sets in between $\rho/\rho^{*}=4$ and $\rho/\rho^{*}=5$, with the ideal A-blocks forming the dense core, as one might expect because of the low or vanishing energy penalty, while the B-blocks form the corona, where their energetically unfavourable overlap is minimized. An unambiguous definition of a micelle remains somewhat arbitrary because of the size polydispersity, but we used the operational convention that all A-blocks whose centres are situated within a distance $r_{c}$ of any other A-block belong to the same cluster. 
\begin{figure}[htb]
\centering
\includegraphics[scale=0.4]{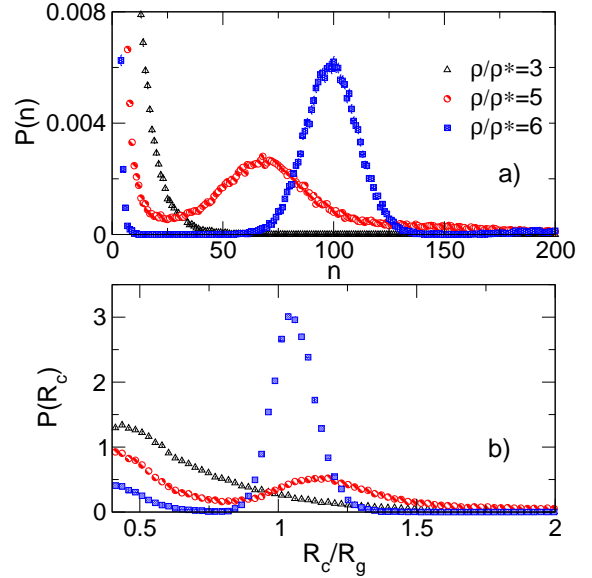}
\caption{(Colour on line) Panel a): probability of a cluster of n monomers. Panel b): probability distribution of the cluster radius $R_{c}$. }
\label{fig:fig2}
\end{figure}
The resulting distributions $P(n)$ of aggregation numbers $n$ are shown in panel $a$ of figure \ref{fig:fig2} for $\rho/\rho^{*}=3, 5$ and $6$, based on a cut-off distance $r_{c}=0.8 R_{g}$. $P(n)$ turns out to be very sensitive to the precise value of $r_{c}$ for $\rho/\rho^{*}\leq 4$ but nearly independent of the cut-off at higher densities, suggesting that well defined micelles appear only beyond an estimated cmc of $\rho/\rho^{*}\sim 5$. At that concentration $P(n)$ exhibits a broad peak around $n\simeq 70$, which narrows and shifts to a mean aggregation number $n\simeq 100$ for $\rho/\rho^{*}\gtrsim 6$. Panel b of figure \ref{fig:fig2} shows the corresponding distribution $P(R_{c})$ of the cluster radii ($R_{c}$) defined as
\begin{equation}
R_{c}^{2}=\frac{1}{n}\sum_{i=1}^{n}\left|\vec{R}_{i}^{A} - \vec{R}_{CM}\right|^{2}
\end{equation}
where $\vec{R}_{i}^{A}$ is the position of the i-th A-block belonging to an n-cluster and $\vec{R}_{CM}$ the position of the instantaneous CM of that cluster. At $\rho/\rho^{*}=3$, $P(R_{c})$ is seen to peak at a low value of $R_{c}$ corresponding to small clusters; the distribution becomes bimodal at $\rho/\rho^{*}=5$, signalling the proximity of the cmc, while at $\rho/\rho^{*}=6$, $P(R_{c})$ shows a sharp peak around $R_{c}\simeq R_{g}$, corresponding to micelles of well defined core size. The peak sharpens and moves to lower values of $R_{c}$ at still larger values of $\rho/\rho^{*}$, indicative of a contraction of the core with copolymer concentration. The internal structure of the micelles is best characterized by the density profiles $\rho_{A}(R)$ and $\rho_{B}(R)$ of the centres of A and B-blocks around the micelle CM. 
\begin{figure}[htb]
\centering
\includegraphics[scale=0.4]{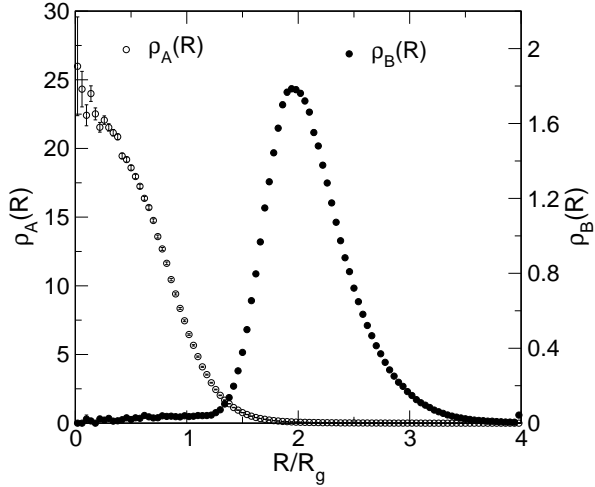}
\caption{$\rho/\rho^{*}=7$. Density profiles of A-blocks (open symbols) and B-blocks (closed symbols) with respect to the CM of each cluster averaged over all clusters with $n\geq 40$. Note the different scales for $\rho_{A}$ (left axis) and $\rho_{B}$ (right axis).}
\label{fig:fig3}
\end{figure}
The profiles shown in figure \ref{fig:fig3} for $\rho/\rho^{*}=7$ are averaged over all micelles formed during the MC simulations, i.e. over size polydispersity. As expected, $\rho_{A}(R)$ peaks at the origin and drops rapidly to zero for $R > R_{g}$, while the profile $\rho_{B}(R)$ practically vanishes for $r<R_{g}$ and peaks at $R\simeq 2R_{g}$, thus defining the corona radius. The sharp distinction between micelle core and corona evident from profiles at $\rho/\rho^{*}=7$ gradually blurs as the copolymer concentration decreases thus signalling that the micelles dissolve into a quasi-homogeneous solution of copolymers.

A periodic sample of 5000 AB copolymers generated about 50 micelles, and the second step of our coarse-graining procedure is to consider the structure of the system of micelles, regarded as well defined spherical colloids. 
\begin{figure}[htb]
\centering
\includegraphics[scale=0.32]{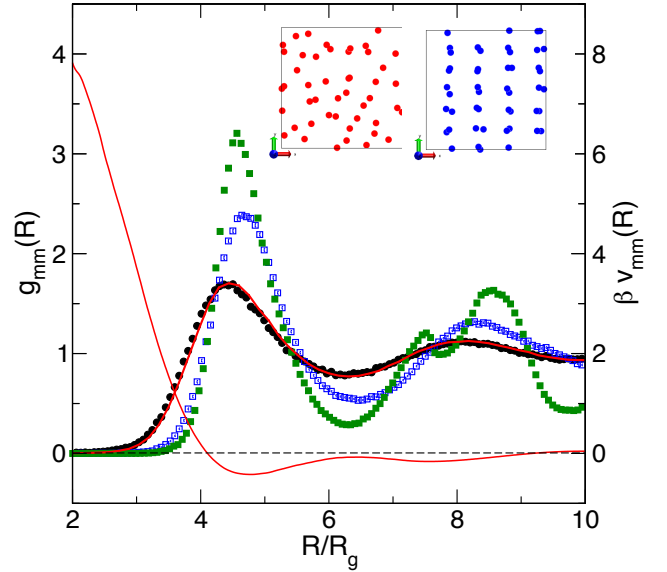}
\caption{(Color on line) Pair distribution function $g_{mm}(R)$ of the CM's of the micelles at $\rho/\rho^{*}=5$ (closed circles), $5.5$ (open squares) and $6$ (closed squares). The HNC potential $v_{m-m}(R)$ at $\rho/\rho^{*}=5$ is shown by a continuous line (in red). Also the corresponding pair correlation is shown as a thick curve in very good agreement with $g_{mm}(R)$ at $\rho/\rho^{*}=5$. Note that the potential scale in on the right axis. Finally, the snapshots in the upper part of the figure show the x-y plane projection of the position of the centre of mass of the micelles in a typical configuration at $\rho/\rho^{*}=5$ (left-disordered) and $6$ (right-ordered).}
\label{fig:fig4}
\end{figure}
Figure \ref{fig:fig4} shows the pair distribution function $g_{mm}(R)$ of the CM's of the micelles, extracted from the MC simulations for copolymers concentrations $\rho/\rho^{*}=5, 5.5$ and $6$. All $g_{mm}(R)$ go rapidly to zero for $R<3R_{g}$, indicative of a strong effective short-range repulsion between micelles. The structure at $\rho/\rho^{*}=5$ is typical of that of a dense colloidal fluid, but it is considerably enhanced at $\rho/\rho^{*}=6$, with the amplitude of the main peak jumping from 1.7 (at $\rho/\rho^{*}=5$) to over 3 while the split second peak in $g_{mm}(R)$ is characteristic of a cubic structure. The sudden change in pair structure is suggestive of a disorder/order transition, and snapshots of micelle configurations, shown in figure \ref{fig:fig4} confirm this interpretation. Note that at $\rho/\rho^{*}=6$, an initially disordered micelle configuration evolves spontaneously to the defective ordered structure shown in figure \ref{fig:fig4} after few tens of thousands MC cycles.

From the MC generated pair distribution function $g_{mm}(R)$ at $\rho/\rho^{*}=5$ one can extract an effective micelle-micelle pair potential using the HNC integral equation of liquid state theory \cite{Hansen86}, according to 
\begin{equation}
v_{mm}(R)=k_{B}T\left\{ g_{mm}(R)-1-c_{mm}(R)-\ln\right[g_{mm}(R)\left]\right\}
\end{equation}
where the direct correlation function $c_{mm}(R)$ is related to $g_{mm}(R)$ by the Ornstein-Zernike relation \cite{Hansen86}. The resulting effective pair potential shown in figure \ref{fig:fig4} exhibits a relatively soft repulsion, which culminates at a finite value at $r=0$, followed by a shallow attractive well. A precise value for the potential at the origin is difficult to obtain because of the noise in $g_{mm}$ at short distance. Assuming a gaussian-like core, a fit to the data in the range $0.5\leq r\leq 3.5$ provides $v_{mm}(0)\simeq 12 k_{B}T$. The accuracy of the inversion procedure is vindicated by a direct MC simulation of micelles (represented as point particles) interacting via this $v_{mm}(R)$: the resulting pair distribution functions, also shown in figure \ref{fig:fig4}, is indistinguishable from the $g_{mm}(R)$ generated in the original simulation of the underlying copolymer system, signalling the success of the second step in our hierarchical coarse-graining procedure. Potentials like $v_{mm}(R)$ shown in figure 4, which remain finite as $R\rightarrow 0$, can give rise to a freezing transition at sufficiently high densities, as shown, e.g., in the case of the gaussian core potential \cite{Stillinger97, Lang00}. The overall scenario described above is also observed when the low density effective potentials $v_{\alpha\beta}(R)$ for the tethered AB copolymer are used (rather than the untethered binary mixture potential, cfg. figure \ref{fig:fig1}), except that the cmc and disorder-order transition are shifted to higher densities ($\rho/\rho^{*}\simeq 6$ and $7$ respectively). Moreover the A-block density profiles (similar to those shown in figure \ref{fig:fig3}), now develop a minimum at $R=0$ at higher densities, so that the micelles tend to become ``hollow'' and increase in size. This is easily understood from the weak short-range repulsion in $v_{AA}(R)$ shown in figure \ref{fig:fig1}.

In summary, we have implemented a two-step coarse-graining procedure for the description of self-assembly and micelle ordering of a simple model for diblock copolymers in a selective solvent. In the first step, the diblock chains are mapped onto ultrasoft dumbells, and the inter and intramolecular potentials are determined by inverting low-density structural MC data for the initial lattice model. MC simulations of large samples of the coarse-grained dumbell model above the overlap density show cluster formation, micellization immediately followed by ordering of the micelles onto a cubic lattice, in qualitative agreement with experimental observation \cite{Lodge05,Hamley,Lodge02,Park05}. In a second step we have determined the effective interaction between the moderately polydisperse spherical micelles, which may be put to good use in a more quantitative investigation of micelle ordering. Our coarse graining methodology raises the general question of the transferability of the low-density effective potentials to semi-dilute copolymer solutions, which we are presently investigating. The present athermal model of a diblock copolymer is clearly an over-simplification, but temperature effects may be easily included by introducing interactions between adjacent monomers on the lattice \cite{Krakoviack03}. The related symmetric ISI model investigated in \cite{Addison05}, where A and B are ideal, but mutually avoiding, exhibits a lamellar phase rather than the present micellar phase, illustrating the strong influence of solvent selectivity. So far we have considered block of equal lengths but the generalisation of our methodology to different relative lengths $f$ of the blocks is straightforward, and is being pursued. Finally we believe that coarse-graining provides new insight into clustering and micellization, as a consequence of the competition between entropy and low energy overlap penalties \cite{Likos01,Einken04,Mladek06}.

C. Addison wishes to thank the EPSRC for a studentship. Part of this work was carried out while C. Pierleoni was a Schlumberger Visiting Fellow at Cambridge, and he wishes to thank SCR for their support. The authors are grateful to A.A. Louis for useful discussions.


\end{document}